\documentclass[aps,twocolumn]{revtex4}

\usepackage{graphicx}% Include figure files
\usepackage{dcolumn}% Align table columns on decimal point
\usepackage{bm}% bold math
\usepackage{amsmath}
\begin{document}

\newcommand{\bk}{{\bf k}}
\newcommand{\bp}{{\bf p}}
\newcommand{\bv}{{\bf v}}
\newcommand{\bq}{{\bf q}}
\newcommand{\bQ}{{\bf Q}}
\newcommand{\br}{{\bf r}}
\newcommand{\bR}{{\bf R}}
\newcommand{\bB}{{\bf B}}
\newcommand{\bA}{{\bf A}}
\newcommand{\bK}{{\bf K}}
\newcommand{\vd}{{v_\Delta}}
\newcommand{\tr}{{\rm Tr}}
\newcommand{\kslash}{\not\!k}
\newcommand{\qslash}{\not\!q}
\newcommand{\pslash}{\not\!p}
\newcommand{\rslash}{\not\!r}
\newcommand{\bs}{{\bar\sigma}}

\title{Coexistence of bulk antiferromagnetic order and superconductivity
in the QED$_3$ theory of copper oxides}

\author{T. Pereg-Barnea and M. Franz}
\affiliation{Department of Physics and Astronomy,
University of British Columbia, Vancouver, BC, Canada V6T 1Z1}

\date{\today}

\begin{abstract}
Within the framework of the QED$_3$ theory of cuprate superconductivity it is 
argued that bulk antiferromagnetic (AF) order can coexist with $d$-wave 
superconductivity in the underdoped region, in agreement with recent 
experiments. The AF order arises from the phase
fluctuating $d$-wave superconductor via the mechanism of spontaneous chiral 
symmetry breaking, provided that fluctuations are sufficiently strong. The 
phase diagram for this coexistence is mapped out by means of analytical and 
numerical solutions of the underlying Dyson-Schwinger equation in the large $N$
limit.
\end{abstract}

\maketitle
%\pacs{74.60.-w,74.60.Ec,74.72.-h}

\begin{figure}[t]
\includegraphics[width=7.6cm]{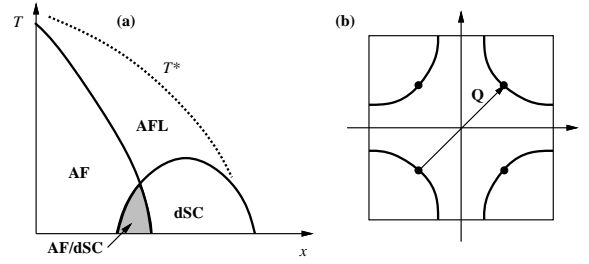}
\caption{\label{fig1}
a) Schematic phase diagram of a cuprate superconductor with AF/$d$SC 
coexistence. AFL denotes algebraic Fermi liquid, the symmetric phase 
of QED$_3$.  b) Cuprate Fermi surface with internodal wavevector ${\bf Q}$.
}
\end{figure}

Interplay between the AF order and superconductivity remains one of the 
central themes in the physics of the high-$T_c$ cuprates 
\cite{anderson1,zhang1,lee1,sachdev1}.
Recently, an appealing new connection between AF and $d$-wave superconducting 
($d$SC) orders has been pioneered, based on the ideas originally articulated 
by Emery and Kivelson \cite{emery1}.
% different from the traditional 
%approaches based on the $t$-$J$ model \cite{chen1}.  
When long range $d$SC order is 
destroyed by thermal or quantum vortex-antivortex fluctuations 
\cite{corson1,ong1,fischer1,balents1,ft1}, 
the resulting state can be either a symmetric
algebraic Fermi liquid \cite{ftv1} or, if the fluctuations are sufficiently
strong, an {\em incommensurate antiferromagnet} \cite{herbut1,ft2,herbut2}
with ordering vector ${\bf Q}$ illustrated in Fig.\ \ref{fig1}.
In the latter case AF order arises through an inherent dynamical instability 
of the underlying effective low energy theory of a phase fluctuating $d$-wave
superconductor, a (2+1) dimensional quantum electrodynamics,
QED$_3$ \cite{ft1,ftv1}. This instability is known as the spontaneous 
chiral symmetry breaking (CSB) \cite{pisarski1,appelquist1,nash1} and is a 
well studied phenomenon in the particle physics.

In this communication we investigate the possibility
that {\em bulk} AF order can set in within the superconducting phase, 
resulting in the region of coexistence of AF and $d$SC orders in the phase 
diagram shown in Fig.\ \ref{fig1}, within the QED$_3$ framework. 
Other approaches to this problem have been considered previously;
see e.g. Ref.\ \cite{chen1}. 
 Recently, experiments have found tantalizing 
hints of such  coexistence in zero applied magnetic 
field in Y and La based cuprates \cite{brewer1,nieder1,sonier1,mook1,sidis1}.
It has been shown previously \cite{fst1} that within the QED$_3$ framework 
such coexistence indeed can occur {\em locally} in the vicinity of fluctuating 
field-induced vortices, as found in neutron scattering \cite{lake1,lake2}, 
$\mu$SR \cite{miller1} and STM \cite{davis1} experiments.

Within the QED$_3$ theory \cite{ft1} the dynamical agent responsible for 
the emergence of AF order is a noncompact U(1) Berry gauge field 
$a_\mu$ which encodes the topological frustration
encountered by the nodal fermions as they propagate on the background of 
the fluctuating vortex-antivortex plasma. In 
the non-superconducting phase Berry gauge
field is {\em massless} \cite{ft1,ftv1,herbut1}. Its quanta, ``berryons'',
have the same effect as photons in ordinary QED$_3$: they mediate long range
interactions between fermions and precipitate the chiral instability 
if the number of fermion species $N$ (pairs of Dirac nodes per unit cell in a
$d$SC) is less than a critical value $N_c\simeq 3.2$
\cite{pisarski1,appelquist1}. In the $d$SC phase, as vortices
bind to finite pairs or loops, the 
Berry gauge field becomes {\em massive} \cite{ft1,ftv1,herbut1}.
Nominaly, one would expect that coupling to such massive gauge field
should become irrelevant for the low energy physics. Here we show
that this is not necessarily so. Based on the approximate analytical and full
numerical solutions of the underlying large-$N$ Dyson-Schwinger equation for
the fermion self energy we find that solutions with broken chiral symmetry 
and finite fermion mass $m_D$ can be found even when the gauge field acquires
small mass $m_a$. This opens up a possibility for the coexistence of bulk 
AF and $d$SC orders in the QED$_3$ theory of underdoped cuprates.

As shown in Refs.\ \cite{ft1,ftv1} the low-energy dynamics of fermionic 
quasiparticles in a $d$-wave superconductor coupled to fluctuating vortices
is governed by the QED$_3$ Euclidean action $S=\int d^3x{\cal L}_D$ with
\begin{equation}
{\cal L}_D\equiv \sum_{l=1}^{N} 
\bar\Psi_l(x) \gamma_\mu (i\partial_\mu -a_\mu)\Psi_l(x) + {\cal L}_B[a(x)].
\label{l1}
\end{equation}
Here, $\Psi_l(x)$ is a four component Dirac spinor representing the 
``topological''fermion excitations associated with a pair of antipodal
nodes, $x=(\tau,{\br})$ denotes the space-time coordinate, and 
$\gamma_\mu$ ($\mu=0,1,2$) are the gamma matrices satisfying
$\{\gamma_\mu,\gamma_\nu\}=2\delta_{\mu\nu}$.  The number  $N$ of 
fermion species is equal to $2n_{\rm CuO}$, with $n_{\rm CuO}$ denoting the
number of CuO planes per unit cell \cite{ftv1}.
Lagrangian ${\cal L}_B$ encodes the dynamics
of the gauge field $a_\mu$ and is given by 
${\cal L}_B[a]={1\over 2}\Pi_{\mu\nu}(q)a_\mu(q)a_\nu(-q)$ with
\begin{equation}
\Pi_{\mu\nu}(q)={1\over e^2}\left(m_a^2+ \alpha|q|+q^2\right)
\left(\delta_{\mu\nu}-\frac{q_\mu q_\nu}{q^2}\right).
\label{ber}
\end{equation}
The mass and Maxwell terms in Eq.\ (\ref{ber}) originate from the bare 
vortex action;
$m_a=0$ in the non-superconducting phase, reflecting the unbound nature of
vortex-antivortex fluctuations, while $m_a>0$ in the superconductor
\cite{ft1,ftv1,herbut1}. The $q$-linear term, with $\alpha=Ne^2/8$, reflects 
the partial screening of the gauge field by the medium of topological 
fermions, expressed to one loop order. 
%The interplay of various terms in 
%Eq.\ (\ref{ber}) will be crucial for determining whether or not CSB occurs.

We analyze the CSB phenomenon in the Lagrangian (\ref{l1},\ref{ber}) 
by solving the
Dyson-Schwinger (DS) 
equation for the fermion self energy $\Sigma(p)$ to leading
order in $1/N$ expansion. This technique is known to be robust against the
higher order $1/N$ corrections even for small $N\sim 2$ \cite{nash1,gusynin1},
and the results appear to be insensitive to vertex corrections \cite{maris1}. 

With the above definitions we can write the DS equation for the 
fermion self energy as \cite{appelquist1} 
\begin{equation}\label{ds1}
\Sigma(p)= { 8 \alpha \over N}\int {d^3k \over (2\pi)^3 }
\frac{\gamma_{\mu}D_{\mu\nu}(p-k)\Sigma(k)\gamma_{\nu}}{k^2+\Sigma^2(k)},
\end{equation}
where $D_{\mu\nu}(q)=\Pi_{\mu\nu}^{-1}(q)$ is the gauge field propagator in 
the Landau gauge.
Following Pisarski \cite{pisarski1}, the simplest way to establish the 
spontaneous mass generation for fermions is to assume that, crudely,  
$\Sigma(p)$ is constant over most of the integration range, $\Sigma(p)=m_D$,
and demand self consistency for the constant value $m_D$:
\begin{equation}
m_D = { 8 \alpha \over N}\int {d^3k \over (2\pi)^3 }
\frac{\gamma_{\mu}D_{\mu\nu}(k)m_D\gamma_{\nu}}{k^2+m_D^2}.
\label{ds2}
\end{equation}
The above DS equation always has a trivial solution 
$m_D=0$, corresponding to the chirally symmetric state. A non-trivial,
symmetry broken solution $m_D>0$ satisfies  
\begin{equation}
1={ 8 \alpha \over \pi^2N}{\int_0^\infty} 
{k^2dk\over(k^2+m_D^2)(m_a^2+\alpha k +k^2)},
\label{ds3}
\end{equation}
where we have taken the trace of both sides and performed the angular 
integrals. For small gauge field mass,
$m_a^2\ll\alpha m_D$, we observe that the fermion mass effectively cuts off 
the integral in the infrared  while $\alpha$ plays the role of the 
ultraviolet cutoff. We may thus approximate Eq.\ (\ref{ds3}) by
\begin{equation}
1={8 \alpha \over \pi^2 N }
\int_{m_D}^{\alpha}{k^2dk \over k^2(\alpha k + m_a^2)}
={8\over\pi^2N}\ln\left({\alpha^2 \over \alpha m_D + m_a^2}\right).
\nonumber
\end{equation} 
In this approximation the fermion mass is given by
\begin{equation}
m_D = \alpha e^{-\pi^2 N/8} - {m_a^2 \over \alpha}.
\label{mass1}
\end{equation}
This is just the classic result of Pisarski \cite{pisarski1} with a small
correction due to the gauge field being massive. Eq. (\ref{mass1}) thus 
suggests that CSB in QED$_3$ could survive in the presence of small gauge 
field mass. Translated into the language of condensed matter physics Eq. 
(\ref{mass1}) suggests that even {\em bound} vortex-antivortex fluctuations 
could give rise to AF instability {\em within} the state with the 
true $d$SC long range order.

One can also perform the integral appearing in Eq.\ (\ref{ds3}) exactly and
confirm that the solution (\ref{mass1}) indeed emerges in the limit 
$m_a^2\ll\alpha m_D$ (up to an unimportant prefactor). The approximation 
employed to obtain Eq.\ (\ref{mass1}) is instructive in that it reveals the
fundamental reason for the insensitivity of CSB to small gauge field mass:
the largest contribution to the RHS of Eq.\ (\ref{ds3}) comes from the region
$m_D<k<\alpha$. Therefore, modifying the gauge boson propagator at momenta
$k\ll m_D$ by introducing the mass term has very little effect on the system.

The above solution is highly suggestive but 
because of the crude approximations involved
it is deficient in at least two ways. First, for $m_a=0$, it fails to 
reproduce the result of a more refined treatment of DS Eq.\ (\ref{ds1}) 
\cite{appelquist1,nash1} that no CSB occurs for $N>N_c$, with $N_c=32/\pi^2$
to leading order in $1/N$ \cite{appelquist1}. Second, Eq.\ (\ref{mass1})
is only valid for $m_a^2\ll\alpha m_D$. In what follows we present a more 
careful treatment of Eq.\ (\ref{ds1}) which is free of the above 
deficiencies. We use these solutions to map out the phase diagram of CSB
as a function of $N$ and $m_a$.

To this end we must relax our assumption of constant $\Sigma(p)$ 
in Eq.\ (\ref{ds1}) and treat 
its full functional dependence on the three-momentum $p$. It is still 
possible to perform the angular integral to obtain
\begin{eqnarray}
\Sigma(p) &=& {4\alpha \over \pi^2 N p}\frac{1}{\sqrt{\alpha^2-4m_a^2}}
\int_0^{\infty} dk \frac{k\Sigma(k)}{k^2+\Sigma^2(k)}
\label{ds4}\\
&\times& 
\left[\theta_1\ln\left( \frac{k+p+\theta_1}{|k-p|+\theta_1} \right) - 
\theta_2\ln\left( \frac{k+p+\theta_2}{|k-p|+\theta_2}\right)\right]
\nonumber
\end{eqnarray}
where $\theta_{1,2} = \frac{1}{2}(\alpha \pm\sqrt{\alpha^2-4m_a^2})$.
In the limit $\alpha^2 \gg m_a^2$ we recover the equation studied in  
Ref.\ \cite{appelquist1},
\begin{equation}
\Sigma(p) = {8 \over \pi^2 N p}\int_{m_a^2 \over \alpha}^{\infty} dk 
\frac{k\Sigma(k)}{k^2+\Sigma^2(k)}\min(k,p),
\label{ds5}
\end{equation}
but with modified lower bound on the integral.
In the massless theory ($m_a=0$), Eq.\ (\ref{ds5}) is solved by linearizing 
the integrand in $\Sigma$ and seeking the solution in the form of a power law,
$\Sigma(p)\sim p^b$. Only solutions with complex $b$ turn out to be
physically admissible placing a condition on the number of fermion species
$N<N_c=32/\pi^2$. The chiral mass is then given by \cite{appelquist1}	
\begin{equation}
m_D\equiv\Sigma(0)=c\alpha e^{-2\pi/\sqrt{N_c/N-1}}, 
\label{mass2}
\end{equation}
with $c$ a numerical constant.

Returning to the theory with massive gauge field, we observe that a simple 
power law ansatz no longer solves the linearized integral equation 
(\ref{ds5}), essentially because of the appearance of the nonzero lower bound
of the integral. Consequently, it is a non-trivial issue to extend 
such an analysis to our case of interest with $m_a > 0$ \cite{gusynin2}.
Fortunately, it is relatively 
straightforward to analyze the full nonlinear, angle integrated, 
DS Eq.\ (\ref{ds4}) numerically. We start from a constant $\Sigma(k)$ and 
numerically evaluate the RHS at $n$ discrete points $p_i$. We then iterate
this procedure until the solution for $\Sigma(p)$ no longer changes 
appreciably between iterations. In this
we adopt an upper cutoff, $\Lambda$, for the integral. We find that when
$\Lambda\gtrsim\alpha$, our solutions no longer depend on $\Lambda$.

\begin{figure}[t]
\includegraphics[angle=270,width=7cm]{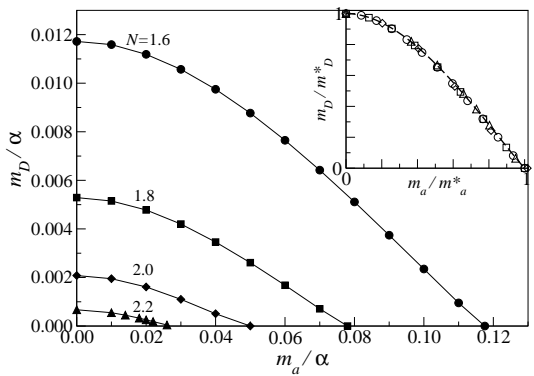}
\caption{\label{fig2}
Numerical solution of Eq.\ (\ref{ds4}) for dynamically generated fermion
mass $m_D$ as a function of gauge field mass $m_a$ for selected values
of $N<N_c$. Inset: curves for various $N$ scaled as described in the text. 
Dashed line represents the solution to Eq.\ (\ref{ds77}).
}
\end{figure}

\begin{figure}[t]
\includegraphics[angle=270,width=6.7cm]{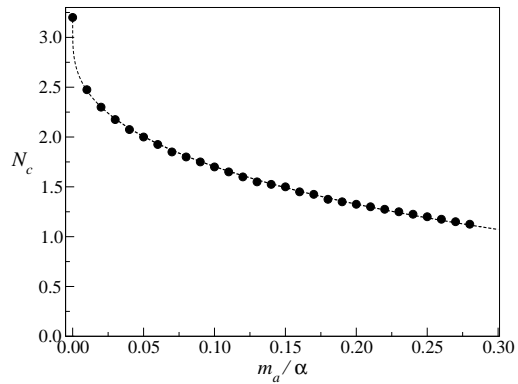}
\caption{\label{fig3}
Numerical solution of Eq.\ (\ref{ds4}) for $N_c$ as a function of 
gauge field mass $m_a$. The dashed line represents a fit to the heuristic 
formula (\ref{nc2}) with ${\cal C}=2.7$.
}
\end{figure}

Figures \ref{fig2} and \ref{fig3} summarize our findings.  
Fig.\ \ref{fig2} displays the dependence of chiral mass $m_D$ on the gauge 
field mass $m_a$ for selected values of $N<N_c$. A notable feature here 
is the surprising resilience of the chiral mass: $m_D$ persists until 
very large values of $m_a$. A hint of why this may be 
 so is offered by a simple
power counting in Eq.\ (\ref{ds3}). This reveals that the relevant scale
actually is not $m_a$ itself but $m_a^2/\alpha$, with $m_a/\alpha\ll 1$.
The latter quantity also enters our estimate for $m_D$, Eq.\ (\ref{mass1}),
which Fig.\ \ref{fig2} confirms to be valid for small $m_a$.

When scaled by $m_D^*$ and $m_a^*$, defined as endpoints of the curves 
in Fig.\ \ref{fig2}, the points for different $N$ fall on the same universal
curve (see inset). One way to understand this universality is from Eq.\
(\ref{ds3}): its {\em exact} solution is given by the implicit equation 
\begin{equation}
f\left({m_D\over m_D^*},{m_a^2\over m_a^{*2}}\right)=0,
\label{ds77}
\end{equation}
with $f(x,y)= x^2\ln x+y^2\ln y+{\pi\over 2}xy$,
independent of $N$. This solution is displayed as a dashed line in the inset
of Fig.\ (\ref{fig2}). We find that the scaling holds very well 
up to $N\approx 2.9$;
closer to $N_c^0\equiv N_c(m_a=0)=32/\pi^2$ 
the agreement gets progressively worse.  
This comparison indicates that the approximation of 
constant self energy, leading to Eq.\ (\ref{ds3}), is very accurate for 
determining $m_D$, as long as $N$ is not too close to $N_c^0$.

Another way of viewing the above data is to consider $N_c$ as a function 
of $m_a$. Fig.\ \ref{fig3} shows this dependence as determined by our 
numerical solution of Eq.\ (\ref{ds4}). As expected $N_c$ is a decreasing 
function of $m_a$. A striking feature is that $N_c$ remains relatively
large for $m_a$ up to significant fraction of $\alpha$. 
%Although we are unable to derive a precise analytical expression for 
The following 
simple argument leads to a highly accurate heuristic formula for $N_c(m_a)$
At low energies there are 
only two scales in the problem: $m_D$ and $m_a$. 
Increasing $m_a$ leads to reduction of $m_D$. It is then natural to expect
that $m_D$ vanishes when the two scales cross, i.e.\ $m_a\approx m_D^*$. Here 
$ m_D^*$ is the fermion mass for given $N$ but with massless gauge field, 
given by Eq.\ (\ref{mass2}). Given our previous insight that the relevant
quantity is $m_a^2/\alpha$ we take our criterion to read 
\begin{equation}
(m_a/\alpha)^2=C(m_D^*/\alpha),		
\label{nc1}
\end{equation}
with $C$ is a numerical constant. Inserting $m_D^*$ from 
Eq.\ (\ref{mass2}) and solving for $N$ we obtain
\begin{equation}
N_c(m_a)={32\over \pi^2}\left[1+\left({\pi/
\ln{{\cal C}\alpha\over m_a}}\right)^2\right]^{-1},
\label{nc2}
\end{equation}
where ${\cal C}=\sqrt{cC}$. As seen from the fit in Fig.\ \ref{fig3} 
for ${\cal C}\approx 2.7$ this 
leads to an excellent agreement with our numerical results. 
We note that the same expression, with ${\cal C}= e^2 \approx 7.4$,
was obtained by 
Gusynin {\it et al.} \cite{gusynin2} from the linearized version of the 
DS equation with an infrared cutoff.

Our results presented above firmly establish the 
existence of nontrivial, chiral
symmetry broken solution of the QED$_3$ Dyson-Schwinger equation in the
presence of a small gauge field mass. The question arises whether this 
represents a true ground state of the system, as is known to be the case
for the massless-photon QED$_3$ \cite{appelquist1}. We believe this remains 
true when $m_a>0$. Imagine turning on the gauge field mass from zero deep
inside the symmetry broken phase, where there is large energy difference 
between
the ground state and the ``false vacuum'' represented by chirally symmetric
phase. It is then clear that introduction of arbitrarily small gauge field
mass cannot raise the energy of the ground state above that of the false 
vacuum: the chirally broken phase must remain a ground state of the system
for some range of values $m_a<m_a^*$, as indicated by our solutions 
(\ref{mass1}) and Fig.\ \ref{fig2}.

Detailed nature of the criticality in the presence of gauge field 
mass appears to be a nontrivial issue. The transition in the ordinary 
massless-photon QED$_3$ is thought to be of a special ``conformal'' variety, 
i.e. a continuous transition of infinite order \cite{appelquist2,miransky1}. 
This is due to the long range nature of interactions mediated by the 
massless gauge field. One would therefore expect that  
in the massive case the transition becomes a conventional second order 
\cite{gusynin2,gusynin3}.

Recently, powerful field theoretic duality 
arguments have been advanced \cite{cohen1} leading to a proposal for an upper 
bound $N_c^0\leq {3\over 2}$. This suggestion finds some support in numerical
simulations of noncompact lattice QED$_3$ \cite{kogut1} which found no 
decisive signal for chiral mass generation at $N=2$. If these results are 
correct this would suggest that the conventional state of the art analysis 
based on the Dyson-Schwinger equation (\ref{ds1}) overestimates $N_c^0$ by
more than a factor of 2. Our
analysis, based on this same technique, would then also be quantitatively
inaccurate. However, we expect its qualitative features to remain valid.  

If we stipulate that chiral symmetry breaking occurs in the massive-photon
QED$_3$, does this necessarily imply coexistence of the AF and $d$SC orders
in cuprates? Answering this question involves additional subtlety 
that is related to the detailed nature of the criticality at the transition 
to the $d$SC state. According to the standard phenomenology 
\cite{ft1,ftv1,herbut1,ft2}, approaching the transition from the above 
(i.e. from the symmetric pseudogap phase) berryon is massless but 
the charge $e$ tends to zero, transforming eventually the Maxwell term 
into the mass term in the $d$SC phase. Since according to 
Eq.\ (\ref{mass2}) $m_D$ is proportional 
to $e^2$ (through $\alpha$) this would suggest that chiral mass should vanish
at the transition. However, there is presumably higher order $(\sim q^4)$
term in the bare berryon action whose prefactor also diverges at the 
transition giving rise to a Maxwell term in the $d$SC phase. The chiral mass
would then be proportional to this new ``charge'' in the $d$SC phase. Our 
calculations above strongly suggest that such a chirally broken phase
can be a globally stable ground state of a system with massive berryons.
Again, while this work might have resolved the nature of the phases on both
sides of the transition, the detailed understanding of the criticality 
itself appears to be a complex problem awaiting future resolution.

QED$_3$ theory as formulated in Refs.\ \cite{ft1,ftv1} therefore appears to
support a locus of AF/$d$SC coexistence in the low doping region of the 
phase diagram Fig.\ \ref{fig1}. In this region the system remains 
superconducting while fermionic excitations become fully gapped. This small 
gap should be observable in thermodynamic and transport measurements. In 
particular the superfluid density should become exponentially activated
in very clean samples at low temperatures.

The authors are indebted to I. F. Herbut, D. E. Sheehy and Z. Te\v{s}anovi\'c
for many penetrating discussions. This work was supported by NSERC; 
in addition M.F. acknowledges the support of the A. P. Sloan Foundation.

\end{document}